\documentclass[11pt]{article}

\usepackage{fullpage}
\usepackage{setspace}
\usepackage{parskip}
\usepackage{titlesec}
\usepackage[section]{placeins}
\usepackage{xcolor}
\usepackage{breakcites}
\usepackage{lineno}
\usepackage{hyphenat}
\usepackage{rotating}
\usepackage{float}

\usepackage[margin=1in]{geometry}
\usepackage{appendix}
\usepackage{amsmath,amsfonts,bm,subfig,enumitem}

\def\vb{{\bm{b}}}

\def\vf{{\bm{f}}}

\def\vr{{\bm{r}}}

\def\vt{{\bm{t}}}

\def\vv{{\bm{v}}}
\def\vw{{\bm{w}}}

\def\vy{{\bm{y}}}
\def\vz{{\bm{z}}}

\def\mA{{\bm{A}}}
\def\mB{{\bm{B}}}

\def\mH{{\bm{H}}}

\usepackage{algpseudocode}
\usepackage{algorithm}

\PassOptionsToPackage{hyphens}{url}
\usepackage[colorlinks = true,
            linkcolor = blue,
            urlcolor  = blue,
            citecolor = blue,
            anchorcolor = blue]{hyperref}
\usepackage{etoolbox}

\usepackage{cite}
\renewenvironment{abstract}
  {{\bfseries\noindent{\abstractname}\par\nobreak}\footnotesize}
  {\bigskip}

\titlespacing{\section}{0pt}{*3}{*1}
\titlespacing{\subsection}{0pt}{*2}{*0.5}
\titlespacing{\subsubsection}{0pt}{*1.5}{0pt}

\usepackage{authblk}

\usepackage{graphicx}
\usepackage[space]{grffile}
\usepackage{latexsym}
\usepackage{textcomp}
\usepackage{longtable}
\usepackage{tabulary}
\usepackage{booktabs,array,multirow}
\usepackage{amsfonts,amsmath,amssymb}
\providecommand\citet{\cite}
\providecommand\citep{\cite}

\newif\iflatexml\latexmlfalse

\AtBeginDocument{\DeclareGraphicsExtensions{.pdf,.PDF,.png,.png,.png,.png,.tif,.TIF,.jpg,.JPG,.jpeg,.JPEG}}

\usepackage[english]{babel}

\usepackage{float}

\iflatexml

\else

\begin{document}

\title{\bf Model-based T1, T2* and Proton Density Mapping Using a Bayesian Approach with Parameter Estimation and Complementary Undersampling Patterns}

\author[1]{Shuai Huang}
\author[2]{James J. Lah}
\author[1,2]{Jason W. Allen}
\author[1]{Deqiang Qiu\thanks{This work is supported by National Institutes of Health under Grants R21AG064405, R01AG072603, R01AG070937 and P30AG066511. Corresponding author: Deqiang Qiu (deqiang.qiu@emory.edu).}
}

\affil[1]{Department of Radiology and Imaging Sciences, Emory University, Atlanta, GA, 30322, USA}
\affil[2]{Department of Neurology, Emory University, Atlanta, GA, 30322, USA}

\vspace{-1em}

 \date{}

\begingroup
\let\center\flushleft
\let\endcenter\endflushleft
\maketitle
\endgroup

Word Count for the body of the text: 4772.

\onehalfspacing

\newpage
\selectlanguage{english}
\begin{abstract}
\normalsize
{\bfseries Purpose:} To achieve automatic hyperparameter estimation for the joint recovery of quantitative MR images, we propose a Bayesian formulation of the reconstruction problem that incorporates the signal model. Additionally, we investigate the use of complementary undersampling patterns to determine optimal undersampling schemes for quantitative MRI.

\par\null

{\bfseries Theory:} 
We introduce a novel nonlinear approximate message passing framework, referred to as ``AMP-PE'', that enables the simultaneous recovery of distribution parameters and quantitative maps.

\par\null

{\bfseries Methods:} We employed the variable flip angle multi-echo (VFA-ME) method to acquire measurements. Both retrospective and prospective undersampling approaches were utilized to obtain Fourier measurements using variable-density and Poisson-disk patterns. Furthermore, we extensively explored various undersampling schemes, incorporating complementary patterns across different flip angles and/or echo times.

\par\null

{\bfseries Results:} AMP-PE adopts a model-based joint recovery strategy, it outperforms the $l_1$-norm minimization approach that follows a decoupled recovery strategy. A comparison with an existing joint-recovery approach further demonstrates the advantageous outcomes of AMP-PE. For quantitative $T_1$ mapping using VFA-ME, employing identical k-space sampling patterns across different echo times produced the best performance. Whereas for $T_2^*$ and proton density mappings, using complementary sampling patterns across different flip angles yielded the best performance.

\par\null

{\bfseries Conclusion:} AMP-PE is equipped with built-in parameter estimation, and works naturally in clinical settings with varying acquisition protocols and scanners. It also achieves improved performance by combining information from the MR signal model and the sparse prior on images.

\end{abstract}

{\bfseries Keywords:} Approximate message passing, Compressive sensing, Quantitative MRI, Hyperparameter estimation, Variable flip angle, Multi-echo, Complementary undersampling pattern, Variable density, Poisson disc.

\sloppy

\pagebreak

\section{Introduction}
\label{sec:intro}
Quantitative MRI (qMRI) techniques are used to measure important tissue parameters, including the $T_1$, $T_2$, and $T_2^*$ relaxation times and proton density. These quantitative maps have proven valuable in detecting subtle changes in tissue properties, and have gained significant traction as potential biomarkers for investigating age-related neurodegenerative diseases  \cite{Eminian:2018, Lee:T1Mapping:2021, Ulla:R2star:2013, Acosta-Cabronero364,Lin:AD:23,Taheri:T1:2011}. Nevertheless, acquiring a fully-sampled dataset in the $k$-space for high-resolution 3D volumetric scans can be time-consuming, leading to patient discomfort and the potential introduction of motion artifacts in reconstructed images. To mitigate this, undersampling techniques are commonly employed to reduce scan time, albeit at the cost of decreased image quality. Consequently, it becomes crucial to incorporate additional prior information into the image reconstruction process to enhance the overall image quality.

First, natural images are widely acknowledged to have sparse representations. Specifically, most of the image wavelet coefficients are close to zero, with the energy concentrated in only a few significant entries. Compressive sensing (CS) leverages this sparsity property and encourages sparse solutions in a suitable basis, such as the wavelet basis \cite{RUP06,CS06,DBWav92}. Various methods, including regularization and Bayesian approaches, have been proposed to enforce the sparse prior on image wavelet coefficients \cite{l1stable06,Yang2010ARO,Baron:2010,Rangan:GAMP:2011,PE_GAMP17,Huang:Entropy:2019}, and we also employ this approach in this paper to enhance image quality.

Second, MR signals are governed by the Bloch equation that includes the underlying tissue parameters and MR acquisition parameters. The inherent correlation among different MR images of the same subject indicates a shared joint sparse and/or low-rank structure, which has been harnessed to achieve improved reconstruction performance \cite{Haldar:JointRecon:2008,Zhao:MRmapping:2015,Trzasko:GroupSparse:2011,Bilgic:JointRecon:2018,Christodoulou:2018,Tamir:T2:2017,Bustin:2019}. Alternatively, the signal model can be directly employed as a prior and integrated into the reconstruction process. This gives rise to a range of model-based joint recovery strategies that offer enhanced performance \cite{Haldar:SuperRes:2009, Block:ModelT2:2009,Sumpf:model:2011,Zhao:model:2014,Zhao:MLE:2016,Akcakaya:Joint:2016,Zhu:diffusion:2017,Guo:qMRI:2017,Wang:T1:2018,Wang:T1:2019,Wang:Physics:2021,Bano:model:2020}. However, a key challenge faced by these joint approaches is the determination of multiple regularization or constraint parameters for optimal results. Manual hyperparameter tuning becomes increasingly cumbersome as the number of images in the model grows, and the process needs to be repeated for various MR scanners or protocols. To tackle this issue, we adopt a Bayesian perspective and employ approximate message passing (AMP) to automatically estimate hyperparameters in this paper.

AMP has gained wide recognition and utilization in sparse signal recovery due to its efficiency and state-of-the-art performance \cite{Donoho:AMP:2009,Baron:2010,Rangan:GAMP:2011,Krzakala:2012:1,Krzakala:2012:2,Vila:EMGAMP:2013,Huang:Quantization:2022,Huang:QSM:2023}. However, the standard AMP was originally developed for linear systems \cite{Donoho:AMP:2009,Rangan:GAMP:2011}, it cannot be used to recover MR tissue parameters in the nonlinear signal model. AMP has been used to recover MR images from linear k-space measurements \cite{Ziniel:DCS:2013,Millard:AMP_MRI:2020,Qiao:AMP_MRI:2020}. Rich et al. later designed a nonlinear AMP framework for phase-contrast MRI and 4D flow imaging \cite{Rich:AMP_PC_MRI:2016,Rich:4Dflow:2018,Aaron:4Dflow:2020}. In this study, we propose an extension to the AMP framework specifically tailored for the nonlinear recovery of MR tissue parameters. In contrast to standard regularization or constraint approaches that necessitate hyperparameter tuning, AMP enables the simultaneous recovery of the signal and hyperparameters in an automatic and adaptive manner. This characteristic renders it an ideal choice for clinical settings involving different scanners and acquisition protocols. We have previously applied this framework to the recovery of $T_2^*$ and phase images \cite{Huang:R2Star:2022}. In this paper, we further extend our approach to jointly recover $T_1$, $T_2^*$, and proton density maps while additionally comparing various undersampling strategies.  

Popular options for undersampling patterns include the variable-density pattern \cite{VD:Cheng:2013}, Poisson-disk pattern \cite{PD:Vasanawala:2011}, among others. To achieve maximum coverage of the k-space, complementary patterns can be adopted at various flip angles and echo times \cite{Haldar:SuperRes:2009, Levine:Complementary:2017, Bliesener:Undersampling:2020, WANG2022118963, Wang:Undersampling:2022, Yang:Complementary:2023}. While previous studies have commonly employed complementary patterns at all flip angles and echo times to enhance multi-contrast MRI, our findings demonstrate that this approach does not hold for quantitative MRI. Specifically, for $T_1$ mapping using VFA-ME, it is advantageous to use identical sampling patterns across different echo times. Conversely, for $T_2^*$ and proton density mappings, employing complementary sampling patterns across different flip angles yields better results. 

\section{Theory}
Variable-flip-angle (VFA) 3D gradient-echo (GRE) has emerged as a popular technique for quantitative MR imaging \cite{Deoni:2003,HAACKE202015}. The acquired measurements span multiple flip angles and echo times. Using the GRE sequence, we acquire undersampled Fourier measurements $\vy_{ij}\in\mathbb{C}^M$ of VFA multi-echo images $\vz_{ij}\in\mathbb{C}^N$ at the $i$-th flip angle $\theta_i$ and $j$-th echo time $t_j$:
\begin{align}
\label{eq:vfa_multi_echo_model}
\vy_{ij}=\mA_{ij}\vz_{ij} + \vw_{ij},
\end{align}
where $i\in\{1,\cdots,I\}$, $j\in\{1,\cdots,J\}$, $\mA_{ij}$ is the measurement matrix, and $\vw_{ij}$ is the measurement noise. The collection of measurements across all flip angles and echo times is denoted by $\vy\in\mathbb{C}^{MIJ}$. The wavelet coefficients $\vv_{ij}$ of the image $\vz_{ij}$ are mostly close to zero, i.e. approximately sparse, in the wavelet domain:
\begin{align}
\label{eq:wavelet_transform}
\vv_{ij} = \mH\vz_{ij}\,,
\end{align}
where $\mH$ is the invertible wavelet transform matrix. The sparse prior on $\vv_{ij}$ is widely used to enhance the quality of reconstructed images \cite{Ye:CS_MRI:2019}. 

Assuming a longitudinal steady-state can be reached with perfect spoiling, the magnitude of the complex MR signal $z_{ij}$ from a spoiled GRE sequence can be expressed as \cite{Bluml:1993}
\begin{align}
\label{eq:signal_model}
\begin{split}
|z_{ij}| &= f_{ij}(z_0,t_1,t_2^*)\\
&=  z_0\cdot\sin\theta_i\cdot\frac{1-\exp(-TR/t_1)}{1-\cos\theta_i\cdot\exp(-TR/t_1)}\cdot\exp(-t_j/t_2^*)\,,
\end{split}
\end{align}
where $z_0$ is the proton density, $t_1$ is the $T_1$ relaxation time, $t_2^*$ is the $T_2^*$ relaxation time, $TR$ is the repetition time.

Utilizing the sparse prior on image wavelet coefficients $\vz_{ij}$, we first calculate the posterior distribution $p_s(\vv_{ij}|\vy)$ of the VFA multi-echo image $\vv_{ij}$ from a probabilistic perspective. Subsequently, we consider this distribution, $p_s(\vv_{ij}|\vy)$, as the prior for the VFA multi-echo image $\vv_{ij}$ and integrate it with the signal model prior for a joint reconstruction of the $T_1$, $T_2^*$, and proton density maps.

\subsection{VFA Multi-echo Image Prior}
We first model the distribution of wavelet coefficients $\vv_{ij}$ using the Laplace distribution. These wavelet coefficients are assumed to be independent and identically distributed (i.i.d.):
\begin{align}
\label{eq:laplace_prior}
    p(v\ |\ \lambda) = \frac{1}{2}\lambda\cdot\exp(-\lambda|v|)\,,
\end{align}
where $\lambda>0$ is the unknown distribution parameter. 

We then model the noise distribution using the complex additive white-Gaussian distribution:
\begin{align}
\label{eq:noise_prior}
    p(w\ |\ \tau_w) = \frac{1}{\pi\tau_w}\exp\left(-\frac{|w|^2}{\tau_w}\right)\,,
\end{align}
where $\tau_w$ is the noise variance.

\begin{figure}[tbp]
\begin{center}
\subfloat[]{
\label{fig:factor_graph_vfa_multi_echo}
\includegraphics[width=.5\textwidth]{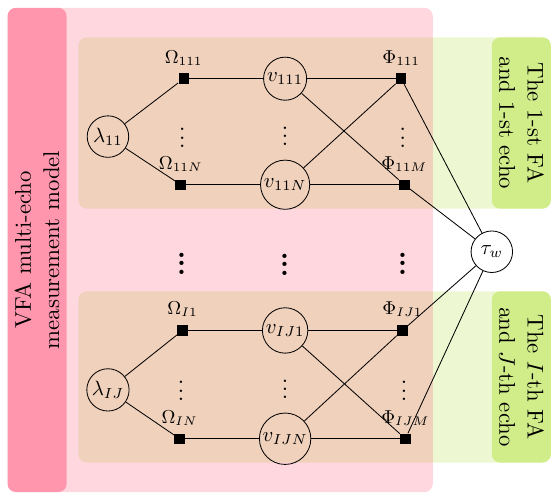}}\\
\subfloat[]{
\label{fig:factor_graph_tissue_parameter}
\includegraphics[width=.8\textwidth]{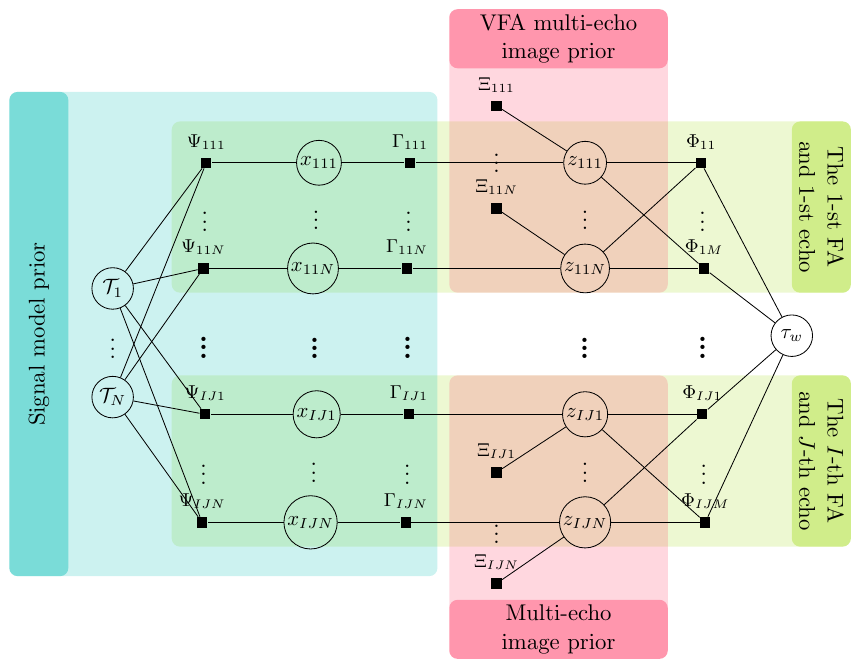}}
\end{center}
\caption{(a) The factor graph used to compute the VFA multi-echo image prior $p_s(\vz_{ijn}|\vy)$ from the measurement model in \eqref{eq:vfa_multi_echo_model}. (b) The factor graph used to recover the tissue parameters $\mathcal{T}_n=\{z_0(n),t_1(n),t_2^*(n)\}$ by combining the VFA multi-echo image prior with the signal model prior in \eqref{eq:signal_model}.}
\label{fig:factor_graph_amp_pe}
\end{figure}

Within the AMP framework, the distribution parameters ${\lambda_{ij},\tau_w}$ are treated as unknown variables \cite{PE_GAMP17}. The factor graph for the forward model \eqref{eq:vfa_multi_echo_model} is illustrated in Fig. \ref{fig:factor_graph_vfa_multi_echo}, where variable nodes are denoted by ``$\bigcirc$'' and contain random variables. The factor nodes, represented by ``$\blacksquare$'', encode the probability distributions of these random variables. In particular, the factor nodes $\Omega_{ijn}$ and $\Phi_{ijm}$ correspond to the signal and noise priors, respectively.
\begin{align}
\Omega_{ijn}\left(v_{ijn},\lambda_{ij}\right) &= p\left(v_{ijn}\ |\ \lambda_{ij}\right)\\
\Phi_{ijm}\left(y_{ijm},\vv_{ij},\tau_w\right) &= p\left(y_{ijm}-\vb_{ijm}\vv_{ij}\ |\ \tau_w\right)\,,
\end{align}
where $\vb_{ijm}$ is the $m$-th row of the measurement matrix $\mB_{ij}=\mA_{ij}\mH^{-1}$. 

Messages about the variable distributions are passed and discussed among the factor nodes until a consensus is reached. As an example, we use the following notations to denote the messages passed between the $n$-th variable node $v_{ijn}$ and the $m$-th factor node $\Phi_{ijm}$ (at the $i$-th flip angle and $j$-th echo):
\begin{itemize}
    \item $\Delta_{v_{ijn}\rightarrow\Phi_{ijm}}$ denotes the message from $v_{ijn}$ to $\Phi_{ijm}$,
    \item $\Delta_{\Phi_{ijm}\rightarrow v_{ijm}}$ denotes the message from $\Phi_{ijm}$ to $v_{ijm}$,
\end{itemize}
where $n\in\{1,\cdots,N\}$ and $m\in\{1,\cdots,M\}$. Both $\Delta_{v_{ijn}\rightarrow\Phi_{ijm}}$ and $\Delta_{\Phi_{ijm}\rightarrow v_{ijn}}$ are functions of the variable $v_{ijn}$, and they are expressed in the ``$\log$'' domain in this paper. The derivation of the AMP algorithm falls beyond the scope of this paper, and algorithmic details can be found in \cite{Rangan:GAMP:2011,PE_GAMP17}. For readers' convenience, a detailed introduction to AMP is provided in Section S-I of the Supporting Information. 

Drawing upon the graphical model theory \cite{Wainwright:Graph:2008}, the posterior distribution of a variable is proportional to the exponential function of the sum of messages passed to that variable:
\begin{align}
p(\lambda_{ij}|\vy)&\propto \exp\left(\sum_n\Delta_{\Psi_{ijn}\rightarrow \lambda_{ij}}\right)\\
p(\tau_w|\vy)&\propto \exp\left(\sum_{ijm}\Delta_{\Phi_{ijm}\rightarrow \tau_w}\right)\\
\label{eq:posterior_p_v}
p(v_{ijn}|\vy)&\propto\exp\left(\Delta_{\Psi_{ijn}\rightarrow v_{ijn}}+\sum_{m}\Delta_{\Phi_{ijm}\rightarrow v_{ijn}}\right)\,.
\end{align}
We can then estimate the distribution parameters $\{\lambda_{ij},\tau_w\}$ using their maximum-a-posteriori (MAP) estimations
\begin{align}
\widehat{\lambda}_{ij} &= \arg\max_{\lambda_{ij}}\ p(\lambda_{ij}|\vy) \\
\widehat{\tau}_w &= \arg\max_{\tau_w}\ p(\tau_w|\vy) \,.
\end{align}

To achieve accurate parameter estimation, it is essential to compute the distributions $p(\lambda_{ij}|\vy)$ and $p(\tau_w|\vy)$ exactly. However, in the AMP framework, the distribution $p(v_{ijn}|\vy)$ can be ``approximated'' by a Gaussian distribution to simplify the calculations without sacrificing accuracy \cite{Minka:2001}:
\begin{align}
\label{eq:posterior_p_v_approx}
p(v_{ijn}|\vy)&\approx\mathcal{CN}\left(v_{ijn}\ \left|\ \mu_{\left(v_{ijn}\right)},\kappa_{\left(v_{ijn}\right)}\right.\right)\,,
\end{align}
where $\mathcal{CN}(\cdot)$ is the complex Gaussian density function, $\mu_{\left(v_{ijn}\right)}$ and $\kappa_{\left(v_{ijn}\right)}$ are the corresponding mean and variance of the wavelet coefficient $v_{ijn}$. Since the wavelet transform $\vv=\mH\vz$ is invertible, we can compute the posterior distribution of the VFA multi-echo image $\vz_{ij}$ from that of the wavelet coefficients $\vv_{ij}$ in \eqref{eq:posterior_p_v_approx} straightforwardly:
\begin{align}
\label{eq:p_VFA_multi_echo}
p_s(z_{ijn}|\vy)\approx\mathcal{CN}\left(z_{ijn}\  \left|\ \mu_{s\left(z_{ijn}\right)},\kappa_{s\left(z_{ijn}\right)} \right.\right)\,,
\end{align}
where $\mu_{s\left(z_{ijn}\right)}$ and $\kappa_{s\left(z_{ijn}\right)}$ are the corresponding mean and variance of the $n$-th image voxel $z_{ijn}$. The distribution $p_s(z_{ijn}|\vy)$ serves as the VFA multi-echo image prior and is combined with the signal model prior in our proposed nonlinear AMP framework.

\subsection{Proposed Nonlinear AMP framework}
The factor graph of the proposed nonlinear AMP framework for reconstructing the tissue parameters is shown in Fig. \ref{fig:factor_graph_tissue_parameter}. We shall introduce a new variable $x_{ijn}$ to represent the MR signal magnitude. It is connected to the complex MR signal $z_{ijn}$ through the factor node $\Gamma_{ijn}$:
\begin{align}
\Gamma_{ijn}(x_{ijn},z_{ijn}) = \delta(x_{ijn}-|z_{ijn}|)\,,
\end{align}
where $\delta(\cdot)$ is the Dirac impulse. The VFA multi-echo image prior and the signal model prior are encoded in the factor nodes $\Xi_{ijn}$ and $\Psi_{ijn}$ respectively:
\begin{align}
\label{eq:Xi_vfa_multi_echo}
\Xi_{ijn}(z_{ijn}) &= p_s(z_{ijn}|\vy)\\
\Psi_{ijn}\left(x_{ijn},z_0(n),t_1(n),t_2^*(n)\right) &= \delta\left(x_{ijn}-f_{ij}(z_0(n),t_1(n),t_2^*(n))\right)\,,
\end{align}
where $f_{ij}(\cdot)$ is the signal model in \eqref{eq:signal_model}.

To simplify the discussion, we narrow our focus to the message passing steps between $\{z_{ijn}\}$ and the tissue parameters $\mathcal{T}_n=\{z_0(n),t_1(n),t_2^*(n)\}$ on the factor graph depicted in Fig. \ref{fig:factor_graph_tissue_parameter}. In this context, we integrate the VFA multi-echo image prior with the signal model prior. The message passing proceeds sequentially through the variable and factor nodes connecting $\{z_{ijn}\}$ and $\mathcal{T}_n$. We then have: 
\begin{enumerate}[label=\arabic*)]
\item {\bfseries Message passing from $\{z_{ijn}\}$ to $\mathcal{T}_n$.} 

The messages proceed as follows:
\begin{align}
\label{eq:msg_z_t}
    \Delta_{z_{ijn}\rightarrow\Gamma_{ijn}} \Rightarrow \Delta_{\Gamma_{ijn}\rightarrow x_{ijn}} \Rightarrow \Delta_{\Gamma_{ijn}\rightarrow x_{ijn}} \Rightarrow \Delta_{x_{ijn}\rightarrow\Psi_{ijn}} \Rightarrow \Delta_{\Psi_{ijn}\rightarrow\mathcal{T}_n}\,.
\end{align}
From the factor graph in Fig. \ref{fig:factor_graph_tissue_parameter}, we can see that the message $\Delta_{z_{ijn}\rightarrow\Gamma_{ijn}}$ is a combination of the messages from $\Phi_{ijm}$ and $\Xi_{ijn}$. Since the distributions of $\{\vv_{ij}, \vz_{ij}\}$ are approximated by Gaussian distributions in AMP \cite{Rangan:GAMP:2011,Minka:2001}, the message $\Delta_{z_{ijn}\rightarrow\Gamma_{ijn}}$ can also be approximated by the logarithm of a Gaussian density function:
\begin{align}
\label{eq:msg_z_gamma}
\begin{split}
\Delta_{z_{ijn}\rightarrow\Gamma_{ijn}} &= \sum_m\Delta_{\Phi_{ijm}\rightarrow z_{ijn}}+\log\Xi_{ijn}(z_{ijn})\\
&= -\frac{1}{\pi\tau_{1(z_{ijn})}}\left|z_{ijn}-\mu_{1(z_{ijn})}\right|^2 + C\,,
\end{split}
\end{align}
where $\mu_{1(z_{ijn})}$, $\tau_{1(z_{ijn})}$ are the corresponding mean and variance of the entry $z_{ijn}$, $C$ is a normalizing constant.

The detailed expressions of the rest messages in \eqref{eq:msg_z_t} are derived in a similar fashion in Appendix \ref{app:sec:message_passing}. Combining the messages from all the factor nodes $\{\Psi_{ijn}\}$ connected to $\mathcal{T}_n$, we can calculate the posterior distribution of the tissue parameters $\mathcal{T}_n$ as follows:
\begin{align}
p(\mathcal{T}_n|\vy) \propto \exp\left(\sum_{ij}\Delta_{\Psi_{ijn}\rightarrow\mathcal{T}_n}\right)
\end{align}
To enhance stability, AMP typically enforces the variances $\left\{\tau_{1(z_{ijn})}\right\}$ in \eqref{eq:msg_z_gamma} to be the same across all the entries in $\vz_{ij}$. The MAP estimations of the tissue parameters are then
\begin{align}
\label{eq:map_tissue_parameter}
\begin{split}
\widehat{\mathcal{T}}_n &= \arg\max_{\mathcal{T}_n}\  p(\mathcal{T}_n|\vy)\\
&=\arg\min_{\mathcal{T}_n}\ \sum_{ij}\left(f_{ij}(z_0(n),t_1(n),t_2^*(n))-\left|\mu_{1(z_{ijn})}\right|\right)^2\,.
\end{split}
\end{align}
The above \eqref{eq:map_tissue_parameter} is a nonlinear least-squares fitting problem, it can be decomposed into three one-dimensional problems with respect to $z_0(n)$, $t_1(n)$, $t_2^*(n)$. The optimization with respect to $z_0(n)$ is convex, and can be solved easily. While the optimizations involving $t_1(n)$ and $t_2^*(n)$ are nonconvex, they can still be efficiently solved using a dictionary-based exhaustive search approach, once the search intervals are properly defined.

\item {\bfseries Message passing from $\mathcal{T}_n$ to $\{z_{ijn}\}$.} 

The messages proceed as follows:
\begin{align}
\label{eq:mse_t_z}
    \Delta_{\Psi_{ijn}\rightarrow x_{ijn}} \Rightarrow \Delta_{x_{ijn}\rightarrow\Gamma_{ijn}} \Rightarrow \Delta_{\Gamma_{ijn}\rightarrow z_{ijn}} \,.
\end{align}
The detailed expressions of the above messages are also given in Appendix \ref{app:sec:message_passing}. Combining the messages from $\Gamma_{ijn}$, $\{\Phi_{ijm}\}$ and the VFA multi-echo image prior $\Xi_{ijn}$, we can finally calculate the posterior distribution of $z_{ijn}$ as follows:
\begin{align}
\begin{split}
p(z_{ijn}|\vy)&\propto\exp\left(\Delta_{\Gamma_{ijn}\rightarrow z_{ijn}}+\log\Xi_{ijn}(z_{ijn})+\sum_m\Delta_{\Phi_{ijm}\rightarrow z_{ijn}}\right)\,.
\end{split}
\end{align}
\end{enumerate}

The rest message passing steps between $z_{ijn}$ and $\tau_w$ are the same as the conventional linear AMP discussed in \cite{Rangan:GAMP:2011} (see Section S-I of the Supporting Information). As mentioned earlier, the message passing process will be performed iteratively until the convergence is reached. The recovered tissue parameters $\widehat{\mathcal{T}}_n$ are given by their MAP estimations in \eqref{eq:map_tissue_parameter}.

\section{Methods}
We collected \emph{in vivo} 3D brain data using a 3T MRI scanner (Prisma model, Siemens Healthcare, Erlangen, Germany), after obtaining written consent from the subjects and receiving approval from the Institutional Review Board of Emory University. The data were acquired using a 32-channel head coil and the GRE sequence. Our objective was to reduce the scan time to approximately 10 minutes, which led us to explore the low-sampling-rate regime, where the undersampling rates varied among ${10\%,\ 15\%,\ 20\%}$. Both retrospective and prospective undersampling schemes were implemented in our experiments. In the retrospective scheme, a fully-sampled dataset was acquired during the scan and then retrospectively undersampled. The reconstructions from the fully-sampled data were used as the ground-truth reference images for comparing different approaches. On the other hand, the prospective scheme involved real-time acquisition of the undersampled dataset. Since it lacked the reference images, its purpose was to validate the feasibility of performing undersampling in a clinical setting.

\begin{figure}[tbp]
\begin{center}
\subfloat[]{
\label{fig:pd_vd_sampling_patterns}
\includegraphics[width=.45\textwidth]{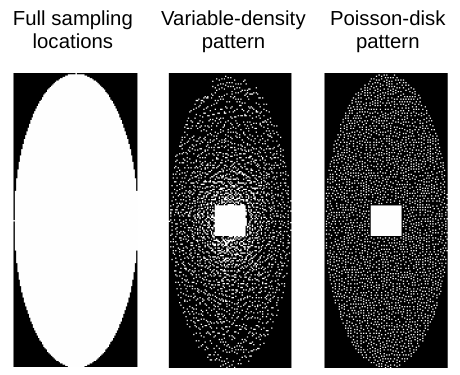}}\\
\subfloat[]{
\label{fig:complementary_sampling_patterns}
\includegraphics[width=.5\textwidth]{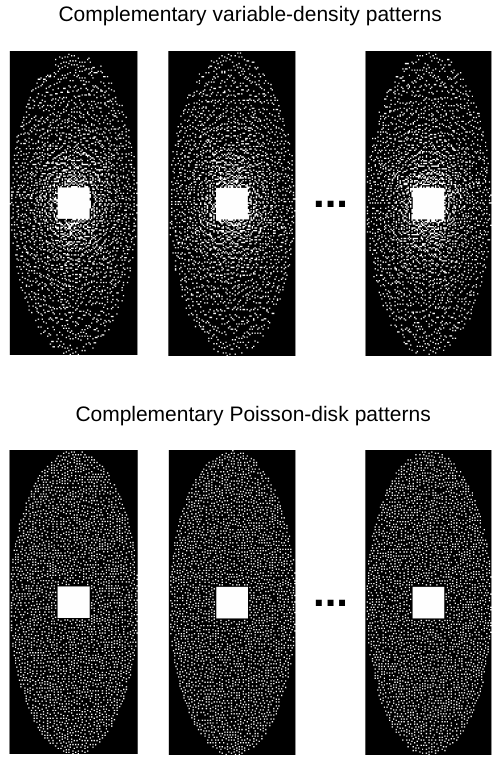}}
\end{center}
\caption{(a) Variable-density and Poisson-disk patterns are two popular undersampling patterns. (b) Complementary undersampling patterns can be adopted across different flip angles and echo times to increase the coverage of $k$-space.}
\label{fig:sampling_patterns}
\end{figure}

\paragraph{Retrospective Undersampling:} The k-space was fully sampled during the scan within an elliptical region of the $y-z$ plane, as illustrated in Fig. \ref{fig:pd_vd_sampling_patterns}. Subsequently, retrospective undersampling was performed in the $y-z$ plane using the undersampling patterns shown in Fig. \ref{fig:pd_vd_sampling_patterns}, while the readout $x$-direction was always fully sampled. For the estimation of sensitivity maps via ESPIRiT \cite{Uecker:ESPIRiT:2014}, the central $24\times 24$ k-space was fully sampled. We employed variable-density and Poisson-disk undersampling patterns and compared their performances. Six subjects, denoted as ``R0--R5'', were recruited for the study. Among them, one subject ``R0'' was randomly chosen as the training dataset (for approaches that required parameter-tuning), while the remaining subjects "R1-R5" served as the test dataset. The acquisition parameters were as follows
\begin{itemize}
\item We included three flip angles = $5$\textdegree, $10$\textdegree, $20$\textdegree; the number of echoes = 4, the first echo time = 7 ms, echo spacing = 8 ms; TR = 36 ms; the number of slices = 96, slice thickness = 1.5 mm; FOV = 256 mm $\times$ 232 mm, in-plane resolution = 1 mm $\times$ 1 mm, bandwidth per pixel = 260 Hz. The acquisition time was 32.83 minutes.
\end{itemize}

\paragraph{Prospective Undersampling:} The prospective protocols were implemented via pulse sequence programming using the ``IDEA'' platform from Siemens. The undersampling took place in the $y-z$ plane in real time, and the readout $x$-direction was always fully sampled. Five subjects, denoted as ``P1--P5'', were recruited for this study. The acquisition parameters were as follows
\begin{itemize}
    \item We included three flip angles = $5$\textdegree, $10$\textdegree, $20$\textdegree; the number of echoes = 4, the first echo time = 7 ms, echo spacing = 8 ms; TR = 36 ms; the number of slices = 96, slice thickness = 1.5 mm; FOV = 256 mm $\times$ 232 mm, in-plane resolution = 1 mm $\times$ 1 mm, bandwidth per pixel = 260 Hz. When the undersampling rates vary in $\{10\%,\ 15\%,\ 20\%\}$, the acquisition times were 5.43, 7.43 and 9.43 minutes respectively. 
\end{itemize}

The double-flip angle method was employed to measure B1 transmit (B1+) field using a echo planar imaging sequence \cite{INSKO199382}. The obtained B1+ field was then combined with Bloch-simulation of the slice-profile of the slab-selective radio-frequency pulse in the 3D GRE sequence to calculate a spatially resolved flip-angle map.

With retrospective undersampling, we investigated the use of complementary undersampling patterns shown in Fig. \ref{fig:complementary_sampling_patterns} for data acquisition through the following undersampling schemes:
\begin{enumerate}[label=\arabic*)]
    \item \emph{U1}: The sampling patterns are complementary across different flip angles and echo times.
    \item \emph{U2}: The sampling patterns are complementary across different flip angles, but the same across different echo times.
    \item \emph{U3}: The sampling patterns are the same across different flip angles, but complementary across different echo times.
    \item \emph{U4}: The sampling patterns are the same across different flip angles and echo times.
\end{enumerate}
After the best undersampling schemes for reconstruction were determined, we applied them in prospective undersampling.

The Daubechies wavelet family was chosen to obtain the sparse representation of an image \cite{DBWav92}. The orthogonal ``db1-db10'' wavelet bases are commonly used, with the complexity of the basis increasing with its order. For the reconstructions of $R_2^*$ map and QSM \cite{Huang:R2Star:2022}, it was observed that employing a higher-order wavelet basis generally resulted in improved image quality. In our experiments, we utilized the db6 basis with 4 levels to strike a balance between wavelet complexity and image quality.

\subsection{Comparison with the Least Squares and $l_1$-norm Regularization Approaches}
\label{subsec:comparison_lsq_l1_amppe}
We first conducted a comparison between the proposed ``AMP with built-in parameter estimation'' (AMP-PE) approach and the baseline least squares (LSQ) approach, as well as the state-of-the-art $l_1$-norm regularization (L1) approach \cite{Ye:CS_MRI:2019}.

\begin{itemize}
    \item The least squares approach:
    \begin{subequations}
    \begin{align}
        &\min_{\vz_{ij}}\ \|\vy_{ij}-\mA_{ij}\vz_{ij}\|_2^2\\
        &\min_{z_0,t_1,t_2^*}\ \sum_{ij}\left(f_{ij}(z_0,t_1,t_2^*)-|z_{ij}|\right)^2\,.
    \end{align}
    \end{subequations}
    The least squares approach does not require parameter tuning, and the solutions can be obtained using gradient descent. Specifically, the recovery of $z_0$, $t_1$, and $t_2^*$ is performed sequentially until convergence. As mentioned earlier, the recovery of $z_0$ is a convex problem and can be easily solved. On the other hand, the recovery of $t_1$ and $t_2^*$ is a nonconvex problem, but it can still be efficiently solved through exhaustive search within the predefined intervals.
    \item The $l_1$-norm regularization approach:
    \begin{subequations}
    \begin{align}
        \label{eq:l1_norm_multi_echo}
        &\min_{\vv_{ij}}\ \|\vy_{ij}-\mB_{ij}\vv_{ij}\|_2^2+\kappa\cdot\|\vv_{ij}\|_1\\
        &\min_{z_0,t_1,t_2^*}\ \sum_{ij}\left(f_{ij}(z_0,t_1,t_2^*)-|z_{ij}|\right)^2\,,
    \end{align}
    \end{subequations}
    where $\mB_{ij} = \mA_{ij}\mH^{-1}$, $\kappa$ is the regularization parameter. The $l_1$-norm of the wavelet coefficients was employed as the regularizer to encourage sparse solutions. The parameter $\kappa=5e-2$ was tuned on the training dataset ``R0'' to achieve optimal performance. We utilized the Fast Iterative Shrinkage-Thresholding Algorithm (FISTA) to solve \eqref{eq:l1_norm_multi_echo} \cite{Beck:FISTA:2009}. The recovery of $z_0$, $t_1$, and $t_2^*$ was also carried out sequentially until convergence.
    \item In the proposed AMP-PE approach, when the undersampling rate is low, the damping operation is necessary to stabilize the AMP update of the wavelet coefficients $\vv$ \cite{Rangan:DampingCvg:2014}. Let $\mu_d^{(t)}(v)$ denote the damped solution in the previous $t$-th iteration, and $\mu^{(t+1)}(v)$ denote the undamped solution in the $(t+1)$-th iteration. The damping operation simply proceeds as follows:
    \begin{align}
    \label{eq:damping}
        \mu_{(z_{ijn})}^{(t+1)}(d) = \mu_{(z_{ijn})}^{(t)}(d)+\alpha\cdot\left(\mu_{(z_{ijn})}^{(t+1)}-\mu_{(z_{ijn})}^{(t)}(d)\right)\,,
    \end{align}
    where $\alpha\in(0,1]$ is the damping rate, $\mu_{(z_{ijn})}^{(t+1)}(d)$ is the damped solution in the $(t+1)$-th iteration. The damping rate $\alpha$ can be regarded as the step size of this iterative update. For a sampling rate of $10\%$, we select $\alpha=0.5$ to slow down the iterative update. However, for relatively higher sampling rates ($\geq 15\%$), we can omit the damping step and choose $\alpha=1$.
\end{itemize}
The L1 approach requires parameter tuning on a training dataset acquired under the same setting as the test data. The LSQ approach, on the other hand, does not require parameter tuning. The AMP-PE approach automatically and adaptively computes the MAP estimations of distribution parameters ${\theta,\tau_w}$. This characteristic makes it a convenient choice for clinical settings across various acquisition protocols and scanners. Unlike the LSQ and L1 approaches, which separate the recovery of VFA multi-echo images from the recovery of tissue parameters, the AMP-PE approach jointly recovers them by combining the VFA multi-echo image prior with the signal prior. Additionally, the maximum numbers of iterations for all three approaches are set to 100.

\subsection{Comparison with the Gradient Support Pursuit Algorithm}
We then proceed to compare AMP-PE with the model-based Gradient Support Pursuit (GraSP) algorithm \cite{Zhao:model:2014}. GraSP, like AMP-PE, also performs a joint recovery of MR tissue parameters, it solves the following constrained maximum likelihood (ML) estimation:
\begin{align}
\begin{split}
\min_{z_0,t_1,t_2^*} \quad &\sum_{ij}\|\vy_{ij}-\mA_{ij}\left(\vf_{ij}(z_0,t_1,t_2^*)\circ\exp(-\boldsymbol i\phi_{ij})\right)\|_2^2\\
\textnormal{subject to} \quad &\|\mH\vz_0\|_0\leq K_{z_0},\,\,\|\mH\vt_1\|_0\leq K_{t_1}\textnormal{ and }\|\mH\vt_2^*\|_0\leq K_{t_2^*}\,,
\end{split}
\end{align}
where $\phi_{ij}$ represents the phase image, ``$\circ$'' denotes the component-wise product, the bold $\boldsymbol i$ symbolizes the imaginary unit, and $K_{z_0}$, $K_{t_1}$, $K_{t_2^*}$ denote the sparsity levels of the proton density, $T_1$, and $T_2^*$ maps, respectively. The sparsity levels were tuned on the training set "R0" and were set to $0.2N$, $0.3N$, and $0.3N$, respectively, to achieve optimal performance. As recommended in \cite{Zhao:model:2014}, the quantitative maps were rescaled to a common range to enhance accuracy and convergence speed. When it comes to the optimization of the quantitative maps $z_0$, $t_1$ and $t_2^*$, GraSP's objective function is formulated as the squared error in relation to the measurements $\vy_{ij}$ in the k-space, whereas the objective functions of LSQ, L1, and AMP-PE are formulated as the squared errors concerning the multi-echo images $f_{ij}$ in the image domain in Section \ref{subsec:comparison_lsq_l1_amppe}. Due to the non-convex nature of all these problems and the difference in the formulation of cost functions, the ground-truth reference image reconstructed from the fully-sampled data for GraSP differs from that of the other approaches. As such, the reconstruction errors of both GraSP and AMP-PE need to be computed relative to their respective reference images. However, the lack of a common reference makes it unfeasible to compare reconstruction errors directly. Thus, the comparison between AMP-PE and GraSP primarily focuses on algorithmic stability, the need for parameter tuning, and computational efficiency.

\section{Results}
\label{sec:results}
\subsection{Comparison with the Least Squares and $l_1$-norm Regularization Approaches}
We first compare different approaches and undersampling schemes using the variable-density pattern. We subsequently highlight the performance differences between the variable-density and Poisson-disk patterns.

\subsubsection{Retrospective Undersampling with the Variable-density Pattern}
Using a brain mask, we computed the normalized root mean square error (NRMSE) within the brain region. Specifically, the reciprocal of the $T_2^*$ map, referred to as the $R_2^*$ map, is frequently employed in brain studies \cite{GHADERY2015925, BARBOSA2015559}. Therefore, we computed the NRMSE with respect to the $R_2^*$ map in this paper. Table \ref{tab:nrmse_t1_r2star_z0} presents the NRMSEs of the recovered $T_1$, $R_2^*$, and proton density maps for subject R1. Due to space constraints, the results for the remaining subjects, R2 to R5, are provided in Tables S1 to S4 in the Supporting Information. Across various sampling rates, both the L1 and AMP-PE approaches generally outperformed the LSQ approach. When the sampling rates were relatively lower at $10\%$ and $15\%$, AMP-PE exhibited superior performance over L1, owing to its joint reconstruction of the tissue parameters. At a relatively higher sampling rate of $20\%$, both AMP-PE and L1 yielded comparable results. The L1 approach requires manual parameter tuning, whereas AMP-PE is equipped with built-in parameter estimation.

\begin{table}[htbp]
\caption{Retrospective undersampling with the variable-density pattern: normalized root mean square errors of recovered $T_1$, $R_2^*$ and proton density $Z_0$ maps from the subject R1 at different sampling rates ($10\%$, $15\%$, $20\%$). Three reconstruction approaches (LSQ, L1, AMP-PE) with four undersampling schemes (U1--U4) are compared in this table.}
\label{tab:nrmse_t1_r2star_z0}
\centering
\begin{tabular}{llcccccccccc}
\toprule
& & \multicolumn{3}{c}{$10\%$} &\multicolumn{3}{c}{$15\%$} &\multicolumn{3}{c}{$20\%$} \\ \cmidrule(lr){3-5} \cmidrule(lr){6-8} \cmidrule(lr){9-11}  
& & LSQ & L1 & AMP-PE & LSQ & L1 & AMP-PE & LSQ & L1 & AMP-PE  \\ \cmidrule(lr){1-2} \cmidrule(lr){3-5} \cmidrule(lr){6-8} \cmidrule(lr){9-11} 

&U1 &0.2235 &0.1974 &0.1901 &0.1784 &0.1545 &0.1530 &0.1560 &0.1358 &0.1362 \\
&U2 &0.2217 &0.1962 &0.1900 &0.1764 &0.1542 &0.1520 &0.1574 &0.1370 &0.1383 \\
&U3 &0.2005 &0.1855 &\bf{0.1769} &0.1667 &0.1498 &\bf{0.1455} &0.1497 &0.1324 &\bf{0.1323} \\
\multirow{-4}{*}{$\hat{\vt}_1$} &U4 &0.2019 &0.1867 &0.1796 &0.1696 &0.1496 &0.1465 &0.1507 &0.1327 &0.1324 \\

\cmidrule(lr){1-2} \cmidrule(lr){3-5} \cmidrule(lr){6-8} \cmidrule(lr){9-11} 

&U1 &0.2098 &0.1953 &0.1687 &0.1355 &0.1190 &0.1159 &0.1094 &\bf{0.0953} &0.0972 \\
&U2 &0.2065 &0.1907 &\bf{0.1664} &0.1352 &0.1191 &\bf{0.1157} &0.1106 &0.0961 &0.0984 \\
&U3 &0.2429 &0.2117 &0.1904 &0.1584 &0.1311 &0.1315 &0.1248 &0.1031 &0.1084 \\
\multirow{-4}{*}{$\hat{\vr}_2^*$} &U4 &0.2457 &0.2140 &0.1919 &0.1597 &0.1324 &0.1337 &0.1253 &0.1038 &0.1093 \\

\cmidrule(lr){1-2} \cmidrule(lr){3-5} \cmidrule(lr){6-8} \cmidrule(lr){9-11}  

&U1 &0.0737 &0.0622 &\bf{0.0604} &0.0538 &\bf{0.0438} &0.0462 &0.0445 &\bf{0.0366} &0.0400 \\
&U2 &0.0753 &0.0630 &0.0617 &0.0549 &0.0445 &0.0470 &0.0451 &0.0371 &0.0407 \\
&U3 &0.0810 &0.0658 &0.0650 &0.0586 &0.0461 &0.0498 &0.0472 &0.0379 &0.0424 \\
\multirow{-4}{*}{$\hat{\vz}_0$} &U4 &0.0827 &0.0668 &0.0665 &0.0601 &0.0472 &0.0512 &0.0481 &0.0384 &0.0430 \\

\bottomrule
\end{tabular}
\end{table}

Different sampling schemes had varying effects on the recovered tissue parameters. Particularly, at low sampling rates of $10\%$ and $15\%$, the performance differences among the schemes became more evident. Regarding $T_1$ mapping, schemes U3 and U4 exhibited comparable performance, surpassing the performance of schemes U1 and U2. Concerning $R_2^*$ mapping, schemes U1 and U2 showed similar performance, significantly outperforming schemes U3 and U4. In the case of proton density mapping, schemes U1 and U2 performed similarly well, outshining schemes U3 and U4.

\begin{figure*}[p]
\includegraphics[width=\textwidth]{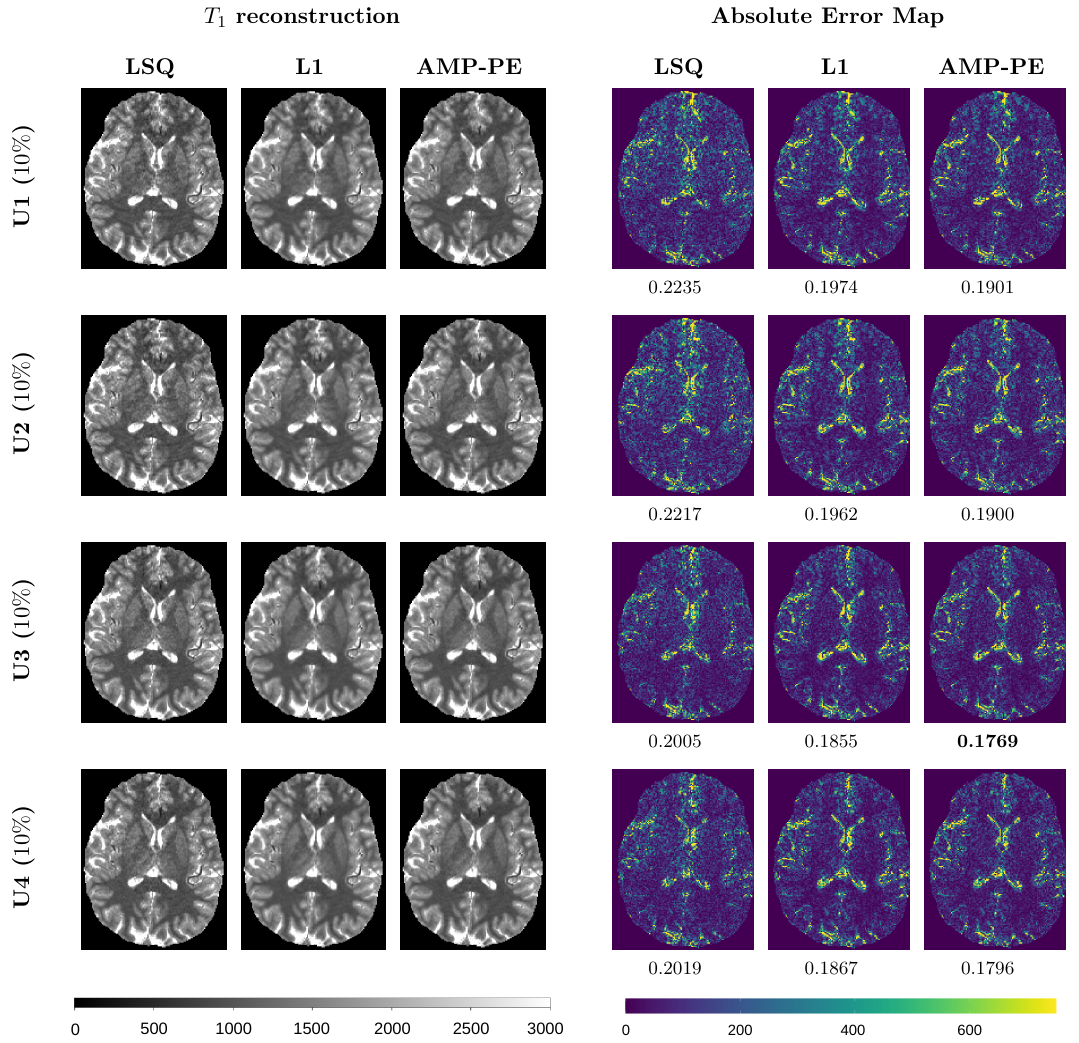}
\caption{Retrospective undersampling with the variable-density pattern at the $10\%$ undersampling rate: recovered $T_1$ maps and absolute error maps from the subject R1 using three reconstruction approaches (LSQ, L1, AMP-PE) and four undersampling schemes (U1--U4). AMP-PE with the U3 scheme achieved the lowest NRMSE of 0.1769.}
\label{fig:compare_reconstruction_t1_10}
\end{figure*}

\begin{figure*}[p]
\includegraphics[width=\textwidth]{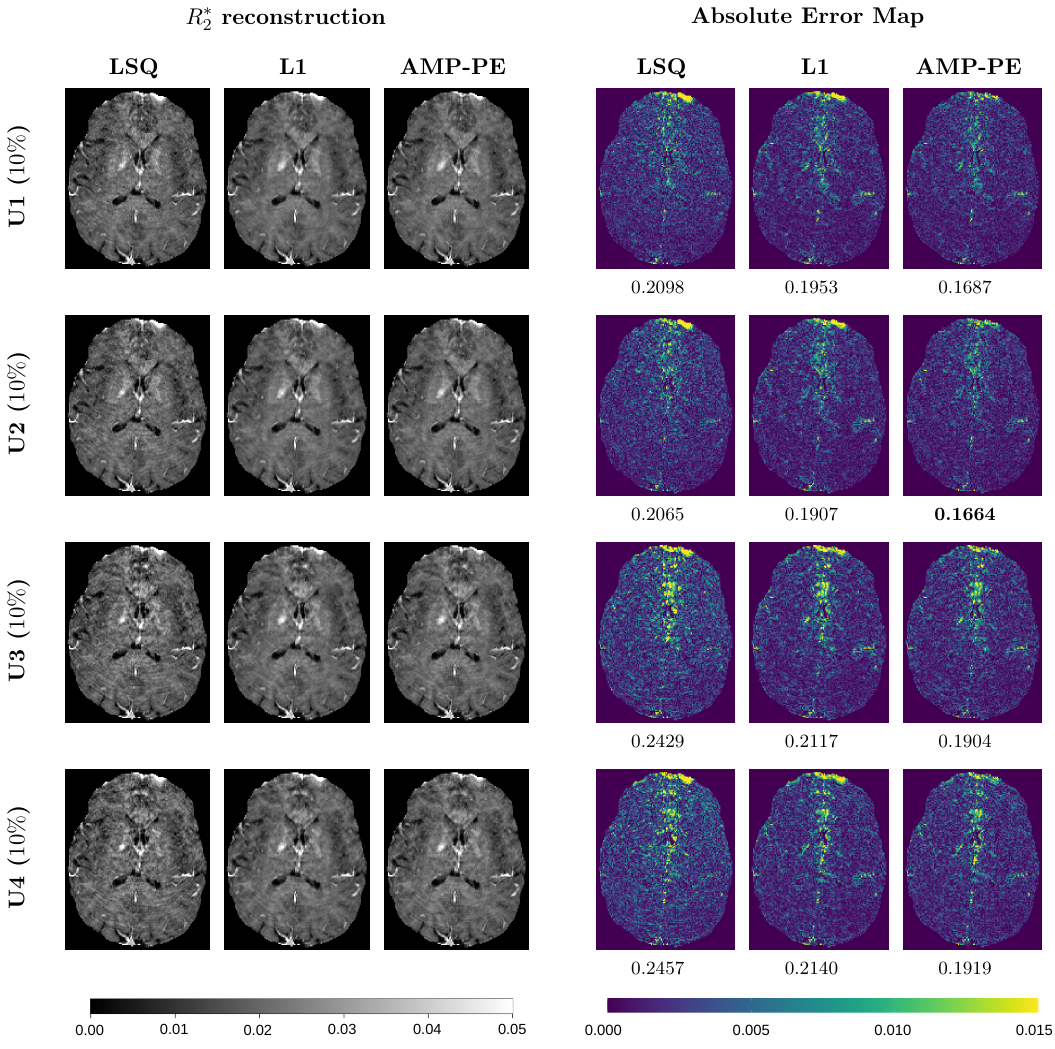}
\caption{Retrospective undersampling with the variable-density pattern at the $10\%$ undersampling rate: recovered $R_2^*$ maps and absolute error maps from the subject R1 using three reconstruction approaches (LSQ, L1, AMP-PE) and four undersampling schemes (U1--U4). AMP-PE with the U2 scheme achieved the lowest NRMSE of 0.1664.}
\label{fig:compare_reconstruction_r2_10}
\end{figure*}

\begin{figure*}[p]
\includegraphics[width=\textwidth]{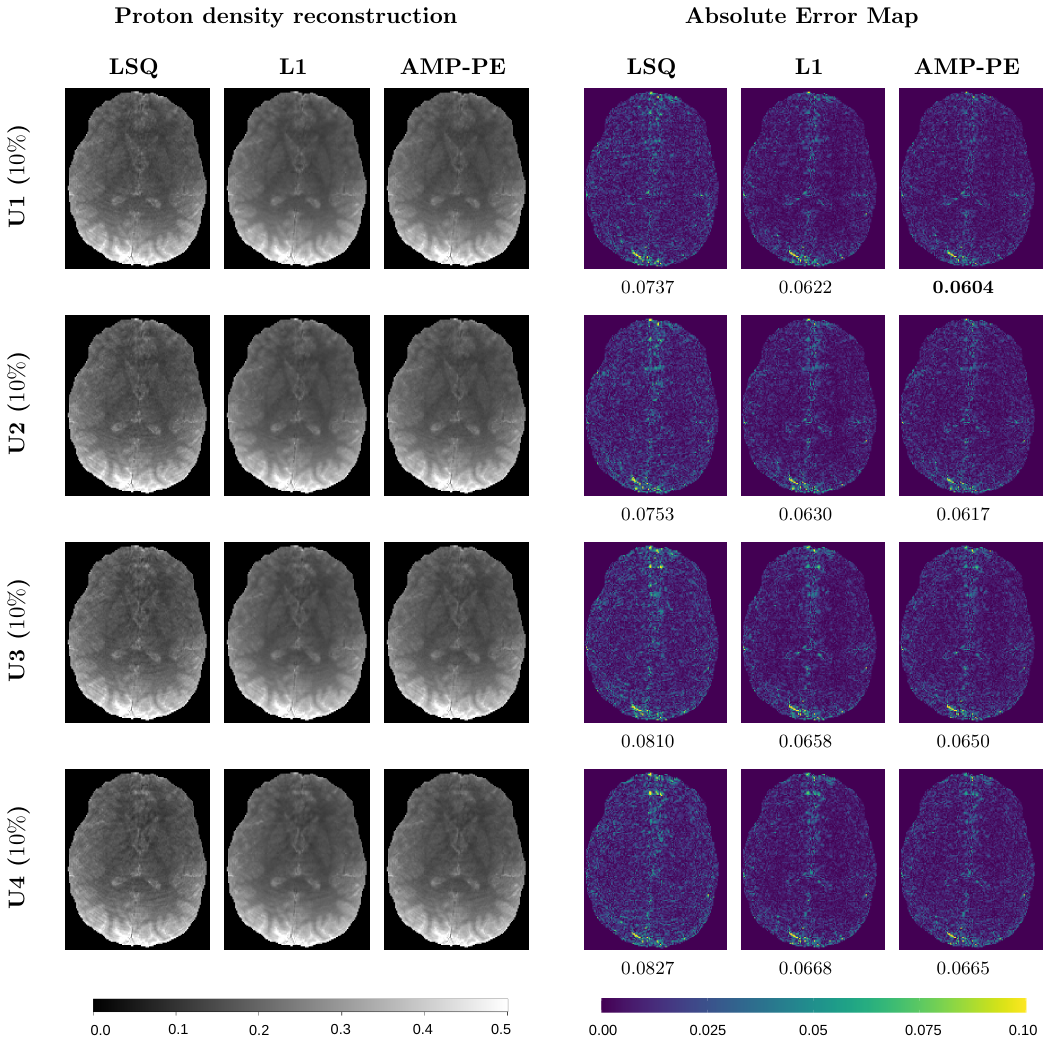}
\caption{Retrospective undersampling with the variable-density pattern at the $10\%$ undersampling rate: recovered proton density maps and absolute error maps from the subject R1 using three reconstruction approaches (LSQ, L1, AMP-PE) and four undersampling schemes (U1--U4). AMP-PE with the U1 scheme achieved the lowest NRMSE of 0.0604.}
\label{fig:compare_reconstruction_x0_10}
\end{figure*}

As an example, we present the recovered images and the corresponding absolute errors for one slice of the 3D brain image from subject R1 at a sampling rate of $10\%$. They are shown in Fig. \ref{fig:compare_reconstruction_t1_10}--Fig. \ref{fig:compare_reconstruction_x0_10}. Due to space constraints, the recovered images at sampling rates of $15\%$ and $20\%$ are provided in Fig. S2--S7 in the Supporting Information. Visual inspection of the images aligns with the quantitative NRMSE findings.

The reconstruction experiments were conducted on the MATLAB platform using a machine (Intel Xeon Gold 5218 Processor, 2.30GHz) with 200 Gb RAM. Using the $15\%$ case as an example, we compared the runtime of each method. The LSQ, L1 and AMP-PE approaches took 1.09, 4.23 and 5.67 hours respectively to complete the reconstruction process.

\subsubsection{Prospective Undersampling with the Variable-density Pattern}
\begin{figure*}[p]
\includegraphics[width=\textwidth]{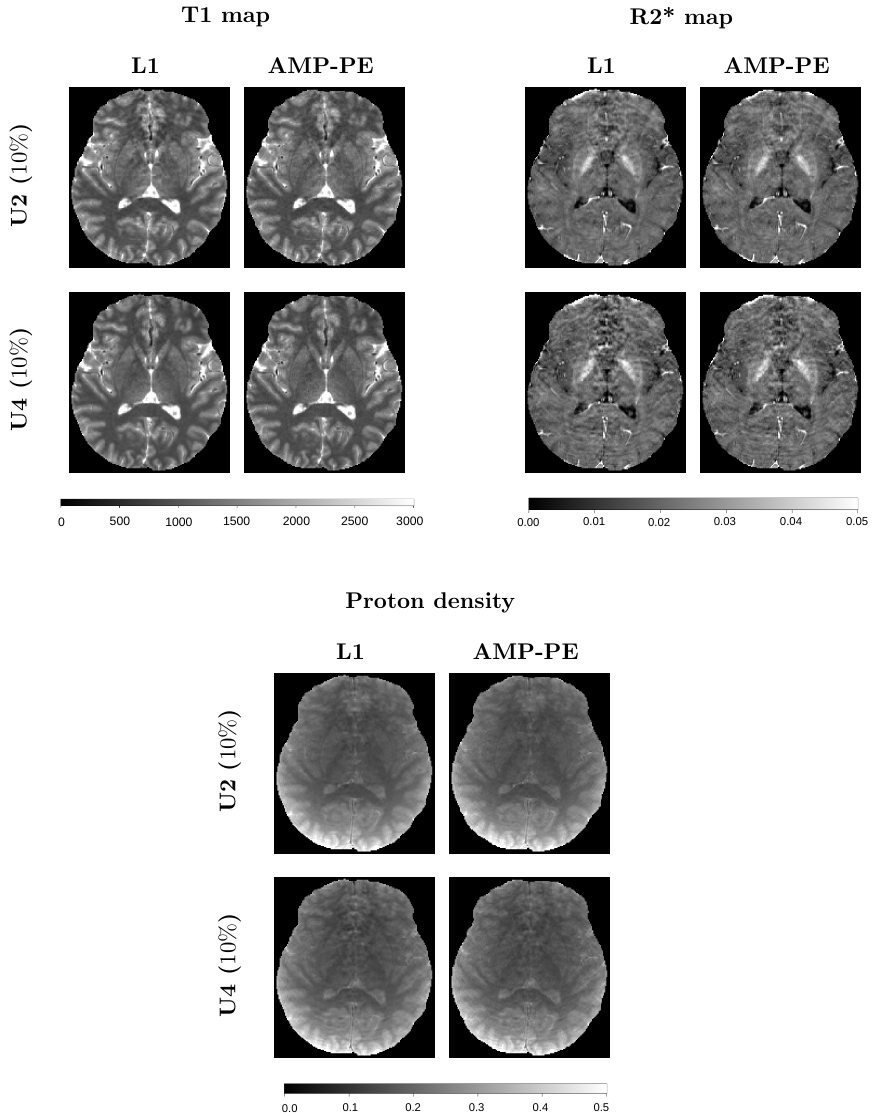}
\caption{Prospective undersampling with the variable-density pattern at the $10\%$ undersampling rate: recovered tissue parameters and absolute error maps from the subject P1 using two reconstruction approaches (L1, AMP-PE) and two undersampling schemes (U2, U4). }
\label{fig:compare_reconstruction_prospective_10}
\end{figure*}

The results obtained from retrospective undersampling revealed that schemes U1 and U2 exhibited similar performance in terms of $R_2^*$ and proton density mapping, while schemes U3 and U4 demonstrated similar performance in $T_1$ mapping. Additionally, U2 and U4 proved to be more straightforward to implement within the acquisition protocol as they did not require different sampling patterns across echo times. Hence, for prospective undersampling in the clinical setting, we opted for U2 and U4. It is important to note that in this scenario, we lacked access to ground-truth reference images needed for calculating NRMSE values. Taking one slice from the 3D brain image from the subject P1 for example, we showcase the recovered images using the L1 and AMP-PE approaches when the sampling rate was $10\%$ in Fig. \ref{fig:compare_reconstruction_prospective_10}. The recovered images for sampling rates of $15\%$ and $20\%$ are provided in Fig. S8--S9 in the Supporting Information. A visual inspection indicates that the prospective undersampling scheme yields comparable and consistent results to the retrospective case.

\clearpage

\subsubsection{Comparison of Variable-density and Poisson-disk Sampling Patterns}
By utilizing the AMP-PE reconstruction approach with U2 and U4 as the undersampling schemes, we can emphasize the performance distinctions between variable-density (VD) and Poisson-disk (PD) patterns, as showcased in Table \ref{tab:nrmse_t1_r2star_z0_vd_pd}. Comprehensive results pertaining to PD, including other approaches and sampling schemes, can be found in Tables S5-S10 of the Supporting Information. Notably, when the sampling rate was $10\%$, VD exhibited significantly superior performance compared to PD. Conversely, at sampling rates of $15\%$ and $20\%$, PD outperformed VD. For instance, considering a selected slice from the 3D brain image of subject P1, the recovered $T_1$, $R_2^*$ and proton density maps are shown in Fig. S10-S12 of the Supporting Information. Visual inspection confirms that VD indeed outperforms PD in the case of $10\%$ sampling rate. However, for the $15\%$ and $20\%$ cases, the superiority of one pattern over the other may vary depending on different brain regions, with PD achieving an overall lower NRMSE.

\begin{table}[tbp]
\caption{Retrospective undersampling: normalized root mean square errors of recovered $T_1$, $R_2^*$ and proton density $Z_0$ maps from the subject R1 at different sampling rates ($10\%$, $15\%$, $20\%$). Using AMP-PE as the reconstruction approach, the variable-density (VD) and Poisson-disk (PD) patterns with selected sampling schemes (U2, U4) are compared in this table.}
\label{tab:nrmse_t1_r2star_z0_vd_pd}
\centering
\begin{tabular}{llcccccccccccc}
\toprule
& & \multicolumn{2}{c}{$10\%$} &\multicolumn{2}{c}{$15\%$} &\multicolumn{2}{c}{$20\%$} \\ \cmidrule(lr){3-4} \cmidrule(lr){5-6} \cmidrule(lr){7-8}  
& & VD & PD & VD & PD & VD & PD  \\ \cmidrule(lr){1-2} \cmidrule(lr){3-4} \cmidrule(lr){5-6} \cmidrule(lr){7-8} 
$\hat{\vt}_1$ &U4 &\bf{0.1796} &0.1939 &0.1465 &\bf{0.1427} &0.1324 &\bf{0.1270} \\ \cmidrule(lr){1-2} \cmidrule(lr){3-4} \cmidrule(lr){5-6} \cmidrule(lr){7-8} 
$\hat{\vr}_2^*$ &U2 &\bf{0.1664} &0.2154 &0.1157 &\bf{0.1133} &0.0984 &\bf{0.0900} \\ \cmidrule(lr){1-2} \cmidrule(lr){3-4} \cmidrule(lr){5-6} \cmidrule(lr){7-8}  
$\hat{\vz}_0$ &U2 &\bf{0.0617} &0.0764 &0.0470 &\bf{0.0448} &0.0407 &\bf{0.0362} \\
\bottomrule
\end{tabular}
\end{table}

\clearpage

\subsection{Comparison with the Gradient Support Pursuit Algorithm}

\begin{figure*}[p]
\includegraphics[width=\textwidth]{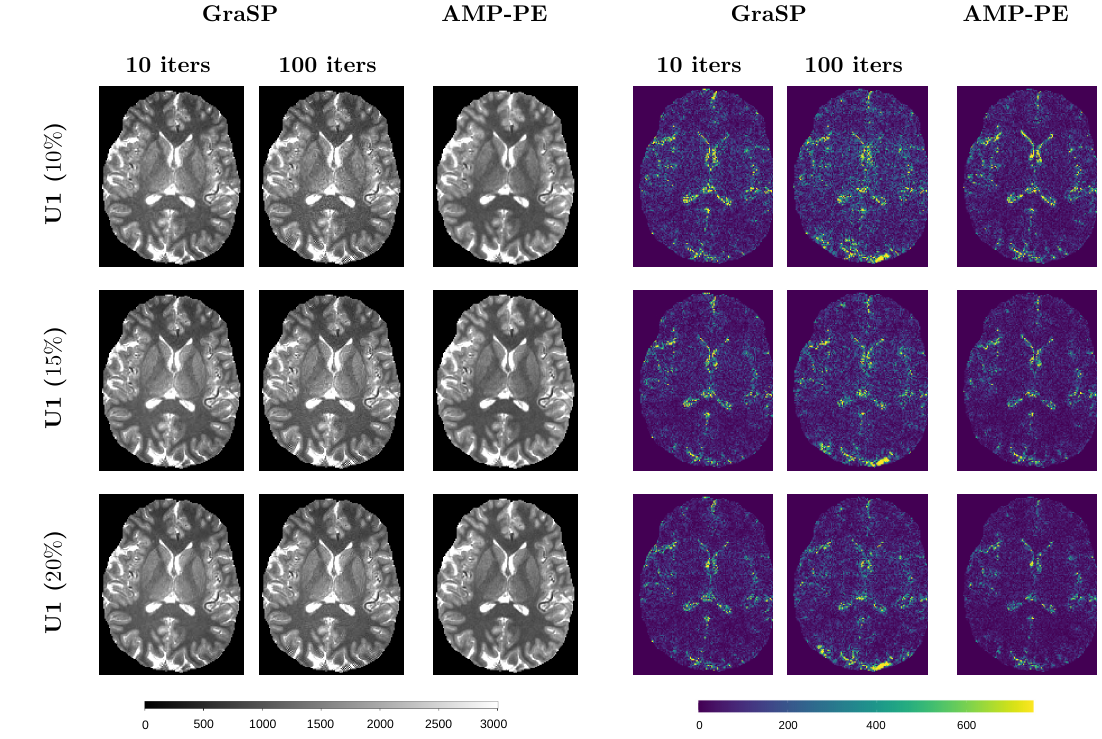}
\caption{Retrospective undersampling with the variable-density pattern: recovered $T_1$ maps and absolute error maps from the subject R1 using GraSP and AMP-PE approaches and U1 undersampling schemes. The recovered images by GraSP with 100 iterations have more errors than those with 10 iterations.}
\label{fig:compare_grasp_amp_t1}
\end{figure*}

We compared GraSP and AMP-PE using the U1 retrospective undersampling scheme with the VD pattern. Taking one slice from the subject R1 for example, we showcase the recovered $T1$ maps and absolute error maps across different sampling rates in Fig. \ref{fig:compare_grasp_amp_t1}. The recovered $R_2^*$ and proton density maps are shown in Fig. S13-S14, and the NRMSEs of recovered images are provided in Table S10 of the Supporting Information. The results showed that GraSP produced more errors compared to AMP-PE. GraSP employs a second-order quasi-Newton method to minimize the objective function \cite{Nocedal:Opt:2006}. However, the use of second-order methods can often result in entrapment within local optima during optimization. Consequently, the images reconstructed by GraSP, with 100 iterations, exhibited more errors in contrast to those with 10 iterations. On the other hand, AMP-PE optimizes the tissue parameters in an alternating fashion. The problem presented in \eqref{eq:map_tissue_parameter} is effectively broken down into three separate one-dimensional problems concerning $z_0$, $t_1$, and $t_2^*$. This approach allows each 1D problem's global optimum to be determined exactly, either through a convex solution or exhaustive search. While the solution achieved using this method is still not guaranteed to be the global optimum of \eqref{eq:map_tissue_parameter}, the results shown in Fig. \ref{fig:compare_grasp_amp_t1} suggest that AMP-PE's methodology proves to be more robust than GraSP's second-order approach. Finally, taking the reconstruction of a 2D slice in the $15\%$ case for example, GraSP (with 10 iterations) and AMP-PE (with 100 iterations) took 4.1 and 4.9 minutes respectively to complete.

\section{Discussion}

Equipped with automatic and adaptive hyperparameter estimation, the proposed nonlinear AMP-PE framework integrates the VFA multi-echo image prior and the signal model prior to jointly recover MR tissue parameters. Notably, we observe that the benefits of joint reconstruction are more pronounced at lower sampling rates, specifically $10\%$ and $15\%$. At the $20\%$ sampling rate, where more data are available, the VFA multi-echo image prior assumes a dominant role, surpassing the significance of the signal model prior. The signal model prior is enforced on the signal magnitudes of each voxel across different flip angles and echo times, representing a voxel-wise local prior. Conversely, the VFA multi-echo image prior originates from the sparse prior on image wavelet coefficients. As the wavelet transform $\mH$ is applied to the entire image, the sparse prior, and therefore the VFA multi-echo image prior, can be considered as global priors in this context.

The convergence of AMP has only been established for linear random Gaussian measurement systems \cite{Rangan:GAMP:2011}. Establishing convergence guarantees for general measurement systems remains an open question. When the sampling rate was as low as $10\%$, we employed the damping operation on the wavelet coefficients $\vv$ in equation \eqref{eq:damping} to ensure the stability of the nonlinear AMP-PE convergence. Additionally, by leveraging a dictionary-based exhaustive search, the nonlinear AMP-PE solves a nonconvex problem to reconstruct tissue parameters in an alternating fashion. It is worth noting that the initialization step plays a crucial role in achieving convergence and avoiding getting stuck in unfavorable local optima. In our study, we discovered that the least-squares solution served as a suitable initialization for AMP-PE.

AMP-PE treats the distribution parameters ${\lambda, \tau_w}$ as variables and is capable of computing their posterior distributions $p(\lambda|\vy)$ and $p(\tau_w|\vy)$ through message passing. However, unlike the posterior distributions of the actual ``image'' variables ${x, z, v}$, the distributions $p(\lambda|\vy)$ and $p(\tau_w|\vy)$ are not approximated as Gaussians in AMP-PE. Therefore, accurately computing the MAP estimations of these parameters becomes challenging since closed-form solutions are typically unavailable. To address this, we relied on a second-order method to compute the MAP estimations of these parameters. To ensure a favorable starting point and avoid undesirable local optima, we initialized the distribution parameters using maximum likelihood estimations based on the least-squares solutions. Additionally, the damping operation can also be applied to the estimated parameters if necessary\footnote{In this study, we did not apply damping to the distribution parameters as the damping applied to the wavelet coefficients had already stabilized AMP-PE for the $10\%$ sampling rate case.}.

Experiments have revealed that both the sampling scheme and sampling pattern in a 3D GRE sequence have an impact on the reconstructed tissue parameters. There is no one-size-fits-all sampling scheme that is suitable for all types of reconstructions. Schemes U1 and U2 are better suited for $T_2^*$ and proton density mapping, while U3 and U4 are more appropriate for $T_1$ mapping. In practical applications, we recommend adopting schemes U2 and U4 for easier implementation in prospective undersampling. As depicted in Fig. \ref{fig:sampling_patterns}, the variable-density (VD) pattern acquires a greater number of low-frequency samples, whereas the Poisson-disk (PD) pattern acquires more high-frequency samples due to its uniform sampling in $k$-space. When the sampling rate is relatively low at $10\%$, the results in Table \ref{tab:nrmse_t1_r2star_z0_vd_pd} demonstrate that having more low-frequency measurements leads to better performance. However, as the sampling rates increase to $15\%$ and $20\%$, high-frequency measurements become more influential, contributing significantly to image quality by capturing more detailed structural information.

\begin{figure}[tbp]
\centering
\includegraphics[width=0.8\textwidth]{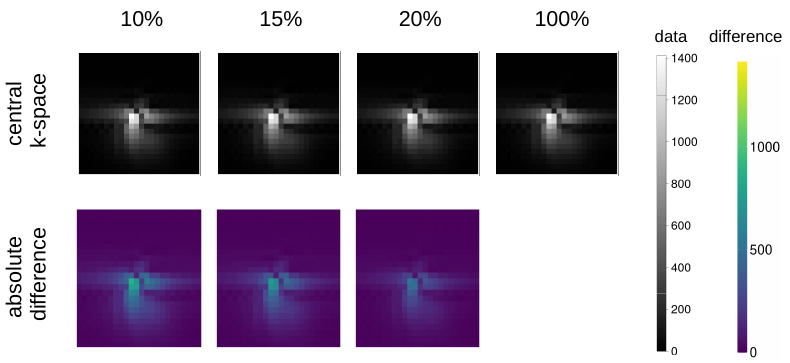}
\caption{Prospective undersampling: the $24\times 24$ central k-space data acquired at different sampling rates ($10\%$, $15\%$, $20\%$, $100\%$). The absolute differences between the undersampled and fully-sampled cases depend on the signal magnitude as well as the sampling rate.}
\label{fig:compare_central_k_space_data}
\end{figure}

In prospective undersampling, we do not have access to ground-truth reference images. Fig. \ref{fig:compare_central_k_space_data} illustrates a comparison of the central k-space data acquired at different sampling rates. The absolute differences in k-space data between the undersampled and fully-sampled cases depend on both the signal magnitude and the sampling rate. Larger signal magnitudes within each undersampled case correspond to larger absolute differences. As the undersampling rate increases, the absolute difference decreases. Consequently, reconstructions based on fully-sampled data cannot serve as reliable ground-truth references in prospective undersampling scenarios, and quantitative evaluations can only be conducted in retrospective undersampling cases.

\section{Conclusion}
We proposed a Bayesian formulation that combines the signal model and sparse prior on VFA multi-echo images to achieve a model-based joint recovery of $T_1$, $T_2^*$, and proton density maps. We designed nonlinear AMP-PE for probabilistic inferences and the reconstruction of quantitative maps. The proposed AMP-PE framework offers automatic and adaptive hyperparameter estimation capabilities, making it a convenient choice for clinical settings with varying acquisition protocols and scanners. The model-based joint recovery approach outperformed the decoupled methods in \emph{in vivo} experiments. Additionally, we explored the use of complementary undersampling patterns in quantitative MRI to further enhance image quality. Our experiments revealed that identical sampling patterns across different echo times are suitable for $T_1$ mapping, while complementary patterns across different flip angles are beneficial for $T_2^*$ and proton density mappings.


\begin{appendices}
\renewcommand\thetable{\thesection\arabic{table}}
\renewcommand\thefigure{\thesection\arabic{figure}}

\section{Message Passing Steps}
\label{app:sec:message_passing}

\begin{enumerate}[label=\arabic*)]
\item {\bfseries Message passing from $\{z_{ijn}\}$ to $\mathcal{T}_n$.} 

The message from $\Gamma_{ijn}$ to $x_{ijn}$ can be calculated as follows:
\begin{align}
\begin{split}
\Delta_{\Gamma_{ijn}\rightarrow x_{ijn}} &= \log \int \Gamma_{ijn}(x_{ijn},z_{ijn})\cdot\exp\left(\Delta_{z_{ijn}\rightarrow\Gamma_{ijn}}\right)\ dz_{ijn}\\
&\approx -\frac{1}{2\pi\tau_{1(z_{ijn})}}\left(x_{ijn}-\left|\mu_{1(z_{ijn})}\right|\right)^2 + C\,.
\end{split}
\end{align}
The message from $x_{ijn}$ to $\Psi_{ijn}$ is
\begin{align}
\Delta_{x_{ijn}\rightarrow\Psi_{ijn}} = \Delta_{\Gamma_{ijn}\rightarrow x_{ijn}}\,.
\end{align}
The message from $\Psi_{ijn}$ to the tissue parameters $\mathcal{T}_n=\{z_0(n),t_1(n),t_2^*(n)\}$ is
\begin{align}
\begin{split}
\Delta_{\Psi_{ijn}\rightarrow\mathcal{T}_n}&=\log\int\Psi_{ijn}\left(x_{ijn},z_0(n),t_1(n),t_2^*(n)\right)\cdot\exp\left(\Delta_{x_{ijn}\rightarrow\Psi_{ijn}}\right)\ dz_{ijn}\\
&=-\frac{1}{2\pi\tau_{1(z_{ijn})}}\left(f_{ij}(z_0(n),t_1(n),t_2^*(n))-\left|\mu_{1(z_{ijn})}\right|\right)^2\,.
\end{split}
\end{align}

\item {\bfseries Message passing from $\mathcal{T}_n$ to $\{z_{ijn}\}$.} 

We can perform the message passing from $\mathcal{T}_n$ to $\{z_{ijn}\}$ in a similar fashion. Specifically, the message from $\Psi_{ijn}$ to $x_{ijn}$ is
\begin{align}
\begin{split}
\Delta_{\Psi_{ijn}\rightarrow x_{ijn}} &= \log\Psi\left(x_{ijn},z_0(n),t_1(n),t_2^*(n)\right)\\
&=\log\delta\left(x_{ijn}-f_{ij}(z_0(n),t_1(n),t_2^*(n))\right)\,.
\end{split}
\end{align}

The message from $x_{ijn}$ to $\Gamma_{ijn}$ is
\begin{align}
\Delta_{x_{ijn}\rightarrow\Gamma_{ijn}} = \Delta_{\Psi_{ijn}\rightarrow x_{ijn}}\,.
\end{align}

The message from $\Gamma_{ijn}$ to $z_{ijn}$ is
\begin{align}
\begin{split}
\Delta_{\Gamma_{ijn}\rightarrow z_{ijn}} &= \log\int\Gamma_{ijn}(x_{ijn},z_{ijn})\cdot\exp\left(\Delta_{x_{ijn}\rightarrow\Gamma_{ijn}}\right)\ dx_{ijn}\\
& = \log\delta\left(|z_{ijn}|-f_{ij}(z_0(n),t_1(n),t_2^*(n))\right)\,.
\end{split}
\end{align}
Combining the messages from $\Gamma_{ijn}$, $\{\Phi_{ijm}\}$ and the VFA multi-echo image prior $\Xi_{ijn}$, we can finally calculate the posterior distribution of $z_{ijn}$ as follows:
\begin{align}
\label{eq:posterior_p_z_final}
\begin{split}
p(z_{ijn}|\vy)&\propto\exp\left(\Delta_{\Gamma_{ijn}\rightarrow z_{ijn}}+\log\Xi_{ijn}(z_{ijn})+\sum_m\Delta_{\Phi_{ijm}\rightarrow z_{ijn}}\right)\\
&\propto\exp\left(-\frac{1}{\pi\tau_{2(z_{ijn})}}\left|z_{ijn}-\mu_{2(z_{ijn})}\right|^2\right)\cdot\exp\left(\sum_m\Delta_{\Phi_{ijm}\rightarrow z_{ijn}}\right)\,,
\end{split}
\end{align}
where $\mu_{2(z_{ijn})}$, $\tau_{2(z_{ijn})}$ are the corresponding mean and variance:
\begin{align}
\mu_{2(z_{ijn})} &= f_{ij}(z_0(n),t_1(n),t_2^*(n))\cdot\frac{\mu_{s(z_{ijn})}}{\left|\mu_{s(z_{ijn})}\right|}\\
\tau_{2(z_{ijn})} &= \tau_{s(z_{ijn})}\,.
\end{align}
The above $\mu_{s(z_{ijn})}$, $\tau_{s(z_{ijn})}$ are the mean and variance of the VFA multi-echo image prior $\Xi_{ijn}(z_{ijn})$ in \eqref{eq:p_VFA_multi_echo},\eqref{eq:Xi_vfa_multi_echo}.

The posterior $p(z_{ijn}|\vy)$ in \eqref{eq:posterior_p_z_final} can be further simplified as
\begin{align}
p(z_{ijn}|\vy) \propto \exp\left(-\frac{1}{\pi\tau_{(z_{ijn})}}\left|z_{ijn}-\mu_{(z_{ijn})}\right|^2\right)\,,
\end{align}
where $\mu_{(z_{ijn})}$, $\tau_{(z_{ijn})}$ are the corresponding mean and variance.
\end{enumerate}

\end{appendices}



\bibliography{ref}

\newpage

{\bfseries\Large Supporting Information}

Additional Supporting Information may be found online in the Supporting Information section.

\paragraph{Supporting Tables S1--S4} Retrospective undersampling with the variable-density pattern: normalized root mean square errors of recovered $T_1$, $R_2^*$ and proton density $Z_0$ maps from the subjects at different sampling rates ($10\%$, $15\%$, $20\%$). Three reconstruction approaches (LSQ, L1, AMP-PE) with four undersampling schemes (U1--U4) are compared in this table.

\paragraph{Supporting Tables S5--S9} Retrospective undersampling with the Poisson-disk pattern: normalized root mean square errors of recovered $T_1$, $R_2^*$ and proton density $Z_0$ maps from the subjects at different sampling rates ($10\%$, $15\%$, $20\%$). Three reconstruction approaches (LSQ, L1, AMP-PE) with four undersampling schemes (U1--U4) are compared in this table.

\paragraph{Supporting Table S10} Retrospective undersampling: normalized root mean square errors of recovered $T_1$, $R_2^*$ and proton density $Z_0$ maps from the subject R1 at different sampling rates ($10\%$, $15\%$, $20\%$). GraSP and AMP-PE were used as the reconstruction approaches under the U1 undersampling scheme with the variable-density (VD) pattern, and the central 10 axial slices were recovered.

\paragraph{Supporting Figure S1} The factor graph of the sparse signal recovery task under the AMP framework: ``$\bigcirc$'' represents the variable node, and ``$\blacksquare$'' represents the factor node.

\paragraph{Supporting Figures S2--S7} Retrospective undersampling with the variable-density pattern at the $15\%$ and $20\%$ undersampling rates: recovered $T_1$, $R_2^*$, proton density maps and absolute error maps from the subject R1 using three reconstruction approaches (LSQ, L1, AMP-PE) and four undersampling schemes (U1--U4).

\paragraph{Supporting Figures S8, S9} Prospective undersampling with the variable-density pattern at the $15\%$ and $20\%$ undersampling rates: recovered tissue parameters and absolute error maps from the subject P1 using two reconstruction approaches (L1, AMP-PE) and two undersampling schemes (U2, U4).

\paragraph{Supporting Figures S10, S11, S12} Retrospective undersampling: recovered $T_1$, $R_2^*$, proton density maps and absolute error maps from the subject R1 using the variable-density (VD) and Poisson-disk (PD) patterns at different sampling rates ($10\%$, $15\%$, $20\%$). AMP-PE and U2 were chosen as the reconstruction approach and undersampling scheme respectively. When the sampling rate is $10\%$, VD achieved a lower NRMSE; when the sampling rates are $15\%$ and $20\%$, PD achieved lower NRMSEs.

\paragraph{Supporting Figures S13, S14} Retrospective undersampling with the variable-density pattern: recovered $R_2^*$, proton density maps and absolute error maps from the subject R1 using GraSP and AMP-PE approaches and U1 undersampling schemes. The recovered images by GraSP with 100 iterations have more errors than those with 10 iterations.

\end{document}